\documentclass[twocolumn,showpacs,preprintnumbers,amsmath,amssymb,superscriptaddress,10pt,aps,prl]{revtex4-2}
\usepackage{epsfig,amsopn}
\usepackage{graphicx}
\usepackage{epstopdf}
\usepackage{sidecap}
\usepackage{amsmath,amssymb,hyperref}
\usepackage{amsthm}
\usepackage{enumerate}
\usepackage{color}

\newcommand{\vect}[1]{\mathbf{#1}}
\newcommand{\ket}[1]{\vert #1 \rangle}

\newcommand{\bracket}[2]{\langle #1 \vert #2 \rangle}

\begin{document}
\title{Surface plasmonics of Weyl semimetals}
\author{Xin Lu}
\affiliation{Laboratoire de Physique des Solides, Univ. Paris-Sud, Universit\'e Paris Saclay, CNRS, UMR 8502, F-91405 Orsay Cedex, France}
\author{Dibya Kanti Mukherjee}
\affiliation{Laboratoire de Physique des Solides, Univ. Paris-Sud, Universit\'e Paris Saclay, CNRS, UMR 8502, F-91405 Orsay Cedex, France}
\author{Mark O. Goerbig}
\affiliation{Laboratoire de Physique des Solides, Univ. Paris-Sud, Universit\'e Paris Saclay, CNRS, UMR 8502, F-91405 Orsay Cedex, France}

\bibliographystyle{apsrev4-2}

\begin{abstract}
Smooth interfaces of topological systems are known to host massive surface states along with the topologically protected chiral one. We show that in Weyl semimetals these massive states, along with the chiral Fermi arc, strongly alter the form of the Fermi-arc plasmon, Most saliently, they yield further collective plasmonic modes that are absent in a conventional interfaces. The plasmon modes are completely anisotropic as a consequence of the underlying anisotropy in the surface model and expected to have a clear-cut experimental signature, e.g. in electron-energy loss spectroscopy.
\end{abstract}

\maketitle
\emph{Introduction.}---%
Weyl semimetals (WSMs) are often considered as a three-dimensional version of graphene since their low-energy $\vect{k} \cdot \vect{p}$ Hamiltonian is described by the massless Weyl equation with linear energy dispersion known as Weyl cones \cite{armitage2018rmp}. WSMs must have an even number of Weyl cones: at least four for inversion symmetry broken WSM and two for time-reversal symmetry broken ones. Previous studies have shown that the bulk dielectric properties such as Friedel oscillations and (magneto-)plasmon in WSMs \cite{dassarme_prl2009,lv_ijmpb2013,panfilov_prb2014,hofmann_prb2015,giri2020} are different from those of graphene \cite{hwang_prb2007,wunsch_njphys2006} due to the increased dimensionality. For example, the bulk plasmon's dispersion in WSMs is gapped and parabolic in momentum while it follows a gapless square-root dispersion in graphene. The chiral anomaly in WSMs \cite{son_prb2013} may be probed by the bulk plasmon \cite{zhou_prb2015} whose dispersion depends on the chirality-resolved chemical potential. As metals, WSMs can also host surface plasmons \cite{ritchie_pr1957} and surface plasmon polaritons \cite{hofmann_prb2016,tamaya_jpcm2019} by using Maxwell's equations in the bulk with a topological Chern-Simons $\theta$-term \cite{zyuzin_prb2012}. In particular, in ferromagnetic WSMs, due to its gapless spectrum and large Berry curvature \cite{kotov_prb2018,pellegrino_prb2015}, electromagnetic waves propagate non-reciprocally, i.e., one direction is preferred. 

Localized surface states can also give birth to original surface plasmons. In WSMs, topologically protected Fermi-arc (FA) states connecting two Weyl cones emerge on the surface due to the bulk-edge correspondence:   the presence of topologically protected edge states  is  dictated  by  the  topological  invariant  of  the twisted bulk band structure. The FA states have been shown to induce a chiral linear FA surface plasmon with total non-reciprocity \cite{song_prb2017,andolina_prb2018,bonaciclosic_jpcm2018,adinehvand_prb2019,chen_prb2019,gorbar_prb2019,ghosh_prb2020}, \textit{i.e.} it propagates only in one direction determined by the the chirality of the FA dispersion. Linear dispersion and total non-reciprocity are highly desirable for further plasmonic applications. Furthermore, a smooth surface of a topological material is known to host massive states called \textit{Volkov-Pankratov} (VP) states \cite{volkov1985two,pankratov1987supersymmetry,tchoumakov2017prbrapid,inhofer2017observation,alspaugh2020volkov,vdberg2020prr_gnr}, along with the protected topological chiral states. These gapped bands can be visualized as pseudo-Landau levels of the system where the smoothness of the interface is modeled as a pseudo-magnetic field \cite{lu2020prb}. Though not protected topologically, their presence may heavily modify the transport \cite{vdberg2020prr_disorder} and magneto-optical properties \cite{lu2019epl,mukherjee2019dynamical} of surfaces of topological materials. 

\begin{center}
\begin{figure}
  \includegraphics[scale=0.65]{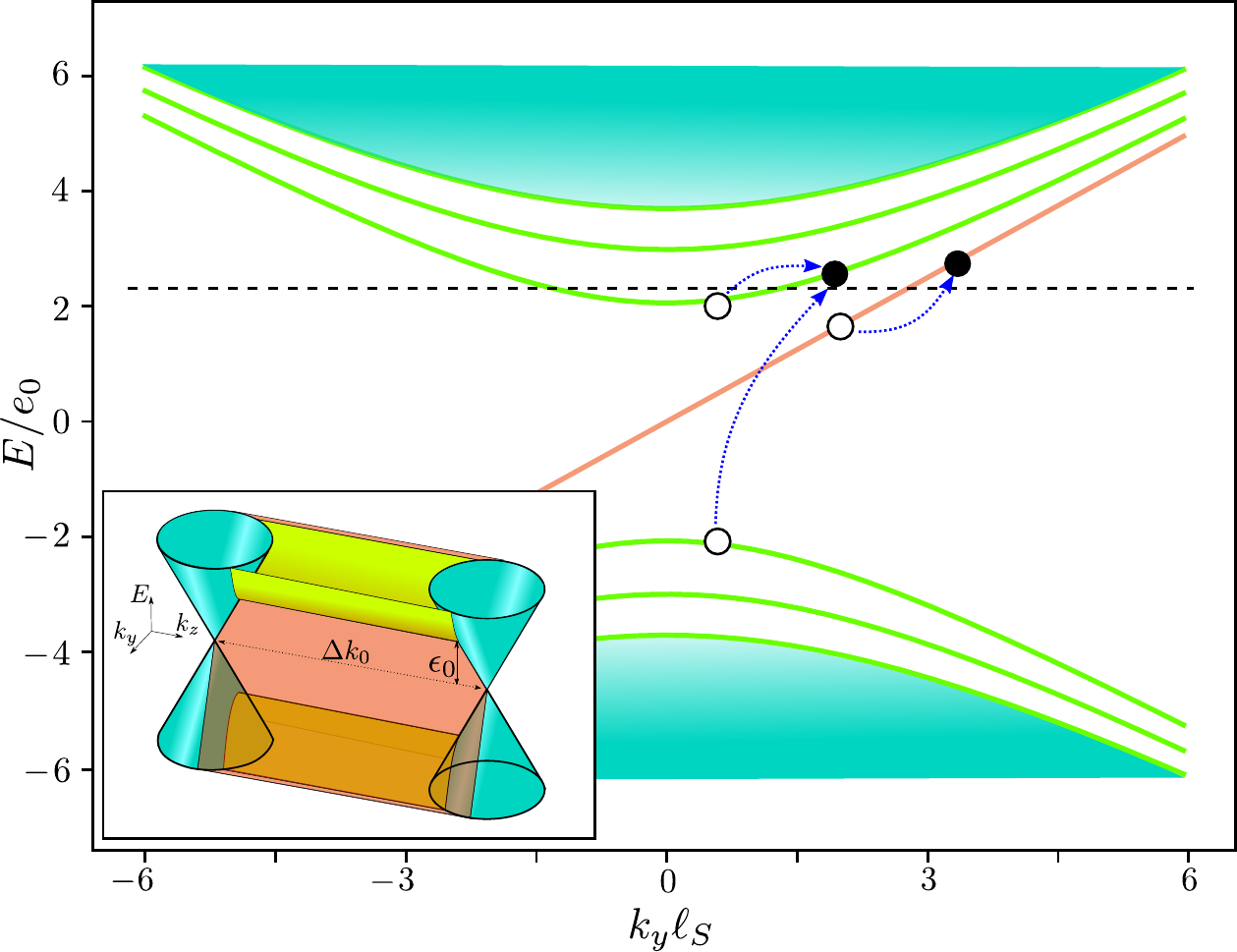}
  \caption{Various optical excitations involving the surface bands for a fixed transverse momentum. The chiral FA is denoted by the red line whereas the massive VP bands are shown in green. As discussed in the main text, for $q_z=0$, only $n\to\pm n$ excitations are allowed. Inset: global view on surface bands connecting two Weyl nodes. Along the transverse momentum, the quasi-1D surface bands do not disperse.}
  \label{Fig:opt_trans}
\end{figure}
\end{center}

Here, we show within  a simple single-boundary model how the FA and the VP states conspire to give rise to new plasmon modes on a smooth surface of a WSM applying \textit{random phase approximations} (RPA). We confirm that the FA plasmon is chiral and exhibits strong anisotropy and a singularity at zero momentum \cite{song_prb2017,andolina_prb2018,adinehvand_prb2019,chen_prb2019,gorbar_prb2019,ghosh_prb2020} because the two-dimensional (2D) dispersion of the FA band evolves into an effectively one-dimensional (1D) one: the energy disperses linearly perpendicular to the $z$-direction connecting two Weyl nodes and remains almost constant along $z$. A spectacular consequence of this anisotropy is the finite gap which the FA plasmon acquires at $q_z=0$ and that vanishes when the longitudinal wavevector $q_z\neq 0$. Moreover, a VP intraband plasmon appears when the chemical potential is above the minimum of the first VP band. Somewhat surprisingly, this plasmon is also non-reciprocal in spite of the $k_y\leftrightarrow -k_y$ symmetric dispersion of the VP bands. As we show below, this is due to coupling to the chiral FA state. We also find a gapped plasmon mode that stems from excitations between two VP bands of same band index and that we call \textit{VP interband plasmon}. 

\emph{Smooth WSM surface}---%
We consider a smooth interface in the $x$-direction between a time-reversal breaking WSM and a trivial insulator modeled by the Hamiltonian \cite{okugawa_prb2014,tchoumakov2017prb,mukherjee2019dynamical}
\begin{align}
\label{eq:hamiltonian}
  H &= v(k_x \sigma_x + k_y \sigma_y) + \left( \frac{k_z^2}{2m} - \Delta + 2\frac{\Delta}{\ell}x \right) \sigma_z ,
\end{align}
where all the material-related parameters are positive and henceforth we use $\hbar = 1$ for notational simplicity. Without the $x$-dependent term, this is the simplest model for a time-reversal breaking two-node WSM with Weyl nodes at $\vect{k} = \eta \sqrt{2m\Delta}\hat{z}$ with opposite chirality $\eta = \pm 1$. The spatially variant gap parameter describes explicitly how the inverted band gap at the center of Brillouin zone is closed and reopened across the interface of width $\ell$ from WSM ($x < 0$) to trivial insulator ($x > \ell$). The smoothness of the surface can be viewed as an effective chiral pseudo-magnetic field $\textbf{B}_P = -\eta 2\Delta /e v \ell \hat{y}$ that couples to the two Weyl nodes of opposite chirality with respective signs. Hamiltonian \eqref{eq:hamiltonian} can be diagonalized by introducing creation and annihilation operators constructed from linear combinations of the $k_x$ and the $x$ dependent terms. Thus, the effective surface bands are reminiscent of Landau levels following the dispersion \cite{supp}
\begin{align}
\label{eq:energy_dispersion_n}
  E_n^\lambda(k_y) = \lambda v \sqrt{k_y^2+\frac{2n}{\ell_S^2}} = \lambda\sqrt{v^2 k_y^2 + n e_0 ^2},
\end{align}
for $n\geq 1$, where $\lambda = \pm$ is the band index and the smoothness of the surface has been encoded in an effective magnetic length $\ell_S = 1/\sqrt{e B_P}$. The VP band gap $e_0=\sqrt{2}v/\ell_S$, which is the separation between the $n=0$ and $n=1$ bands at $\textbf{k}=0$, sets the characteristic energy scale of this surface model. The FA is naturally described here by the $n=0$ band with 
\begin{align}
\label{eq:energy_dispersion_0}
  E_0(k_y) = v k_y
\end{align}
and breaks the symmetry $k_y \to -k_y$, its counterpart with opposite sign of the dispersion being localized at the other surface of the WSM that we do not consider here. The FA state is independent of the surface details such as its smoothness, \textit{i.e.} the band dispersion does not depend on $\ell$, indicating its topological nature. However, the $n\geq 1$ VP bands depend strongly on the surface modeling. In the sharp-surface limit ($\ell \to 0$), the VP bands rise up in energy and eventually merge with the bulk states when $v\sqrt{2n}/\ell_S \sim \Delta$, while only the FA state survives.

From Eq. \eqref{eq:energy_dispersion_n}, we can see that the VP bands are completely flat in the $k_z$-direction until they hybridize with the bulk Weyl cones, as shown in the inset of Fig. \ref{Fig:opt_trans}. In spite of being embedded in a 2D $(k_y,k_z)$ manifold, the VP bands are effectively 1D and thus exhibit van Hove singularities in the density of states at the band extrema. The underlying 2D nature and the $k_z$-dependence is encoded in the location of the surface states: along the interface, they have a Gaussian profile of a characteristic width $\ell_S$ centered at $\langle x\rangle = B_P(\Delta-k_z^2/2m)$. This, as shown below, results in non-diagonal overlap matrix elements for excitations in the $k_z$-direction.

\emph{Quasi-two-dimensional RPA.}-- In order to analyze the behavior of surface electrons, consider the non-interacting dynamical polarization
\begin{align}
    \chi^{(0)}_{i,j}(\vect{q},\omega) =& \sum_{i,j} \frac{1}{V} \sum_{\vect{k}} \frac{f_D (E_i (\vect{k}) ) - f_D (E_j (\vect{k+q}) )}{\omega + E_i (\vect{k}) - E_j (\vect{k+q}) + i \delta} \nonumber \\ & \times |F_{i,j} (\vect{k},\vect{k+q})|^2,
\end{align}
where $\delta= 0^{+}$ and the $i,j$ indices are shorthand notations for both band labels $n$ and $\lambda$. In general, the overlap matrix element $F_{i,j}$ is not diagonal because of the abovementioned $k_z$-dependence of the eigenstates so that $\chi^{(0)}_{i,j}$ is generally a tensor. However, for $q_z=0$, the particle-hole excitations are also 1D and $F_{i,j}$ becomes diagonal, meaning that only excitations from $n$ to $\pm n$ are possible. The RPA dielectric function then retrieves its usual form
\begin{align}
\label{eq:rpa}
    \epsilon^{\text{RPA}}(q_y,\omega) = 1- V_{\text{2D}}(q_y) \chi^{(0)}(q_y,\omega),
\end{align}
where $V_{\text{2D}}(q_y) = e^2/2 \epsilon_0\epsilon_r |q_y|$, and $ \chi^{(0)}$ is the non-interacting charge susceptibility, in terms of the environmental dielectric constant $\epsilon_r$ \cite{supp}.

When $q_z \ne 0 $, the overlap matrix element $F_{i,j}(\vect{k},\vect{k}+\vect{q})$ is more involved. However, in the long-wavelength limit, the off-diagonal term $F_{i,j}$ is proportional to $q_z^{|n_i - n_j|}$ \cite{supp} so that the $n \to \pm n$ excitations still remain the leading contributions to the charge susceptibility. Nevertheless, due to the complicated form of the off-diagonal terms, we cannot factorize the Coulomb interaction operator \cite{supp} even in the long-wavelength limit when several VP bands are present. We therefore consider only the chiral FA and the $n =\pm 1$ VP bands (\textit{threeband model}), where \cite{supp}
\begin{align}
\label{eq:chi_qz_nonzero}
\chi^{(0)}(\vect{q},\omega) = \sum_{i,j} \chi^{(0)}_{i,j}(\vect{q},\omega),
\end{align}
and $\chi^{(0)}_{i,j}$ are the contributions by the excitation $ (n_i, \lambda_i) \to (n_j, \lambda_j)$. Accordingly, we can generalize $(q_y,\omega)$ in Eq. \eqref{eq:rpa} to $(\vect{q},\omega)$.

\begin{center}
\begin{figure}
  \includegraphics[width=0.48\textwidth]{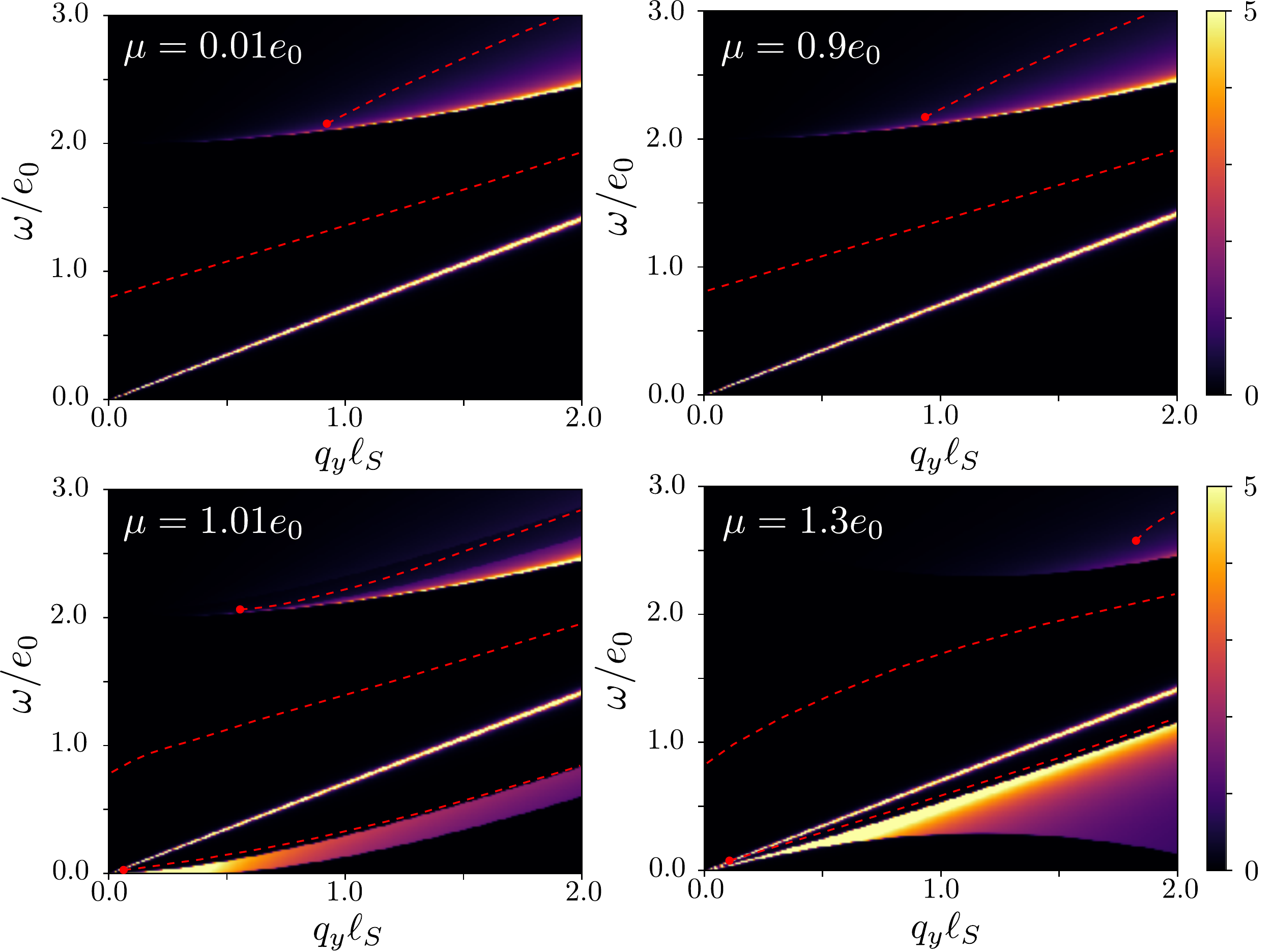}
  \caption{Profile of the imaginary part of the non-interacting dynamical polarization $-\text{Im}(\chi^{(0)})$ in the $(q_y,\omega)$ phase space at $\mu = 0.01 e_0, 0.90 e_0, 1.01 e_0$ and $1.30 e_0$ for $q_z = 0$. The zeros of the real part of $\epsilon^{\text{RPA}}$ (red dashed lines) indicate the plasmon modes.}
  \label{Fig:qz=0}
\end{figure}
\end{center}

\begin{center}
\begin{figure}
  \includegraphics[width=0.48\textwidth]{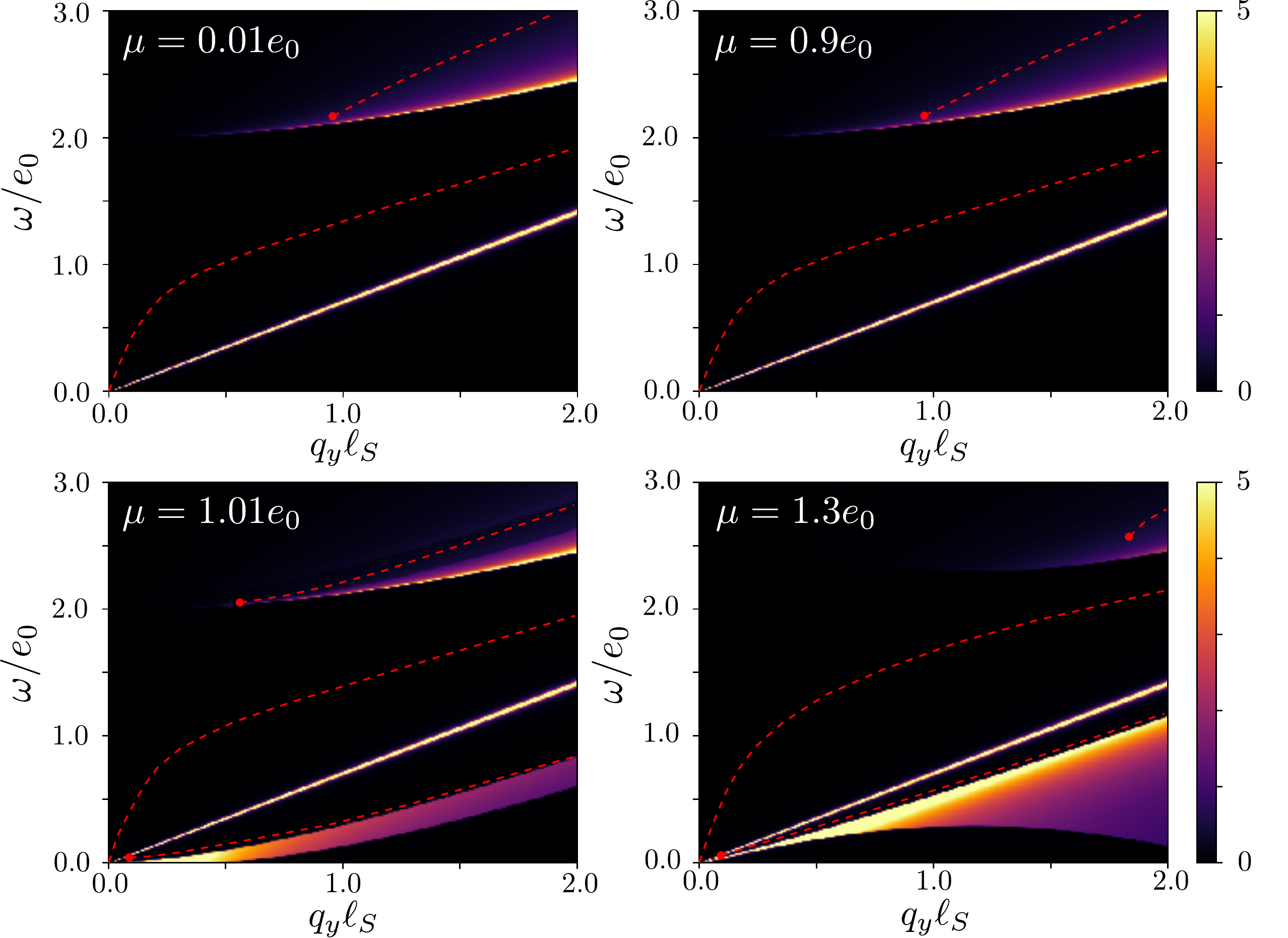}
  \caption{Same as Fig. \ref{Fig:qz=0} for $q_z\ell_S = 0.2$. The zeros of the real part of $\epsilon^{\text{RPA}}$ (red dashed lines)
  indicate the plasmon modes.}
  \label{Fig:qz_nonzero}
\end{figure}
\end{center}

\emph{Plasmons}---%
We summarize our results for $q_z=0$ in Fig. \ref{Fig:qz=0} and for $q_z\ell_S = 0.2$ in Fig. \ref{Fig:qz_nonzero} where we numerically calculate the profile of $-\text{Im}(\chi^{(0)})$ in the $(q_y>0,\omega>0)$-plane, for different values of the chemical potential $\mu$ and a given disorder amplitude $\delta = 0.01$. Indeed, the imaginary part of $\chi^{(0)}$ bears important information about possible electronic excitations and therefore damping of the plasmon modes, indicated by the red dashed lines (zeros of the real part of $\epsilon^{RPA}$). The plasmon modes are only long-lived and undamped in the black regions where $\text{Im}(\chi^{(0)})=0$. Within the abovementioned threeband model, one obtains three particle-hole continua, with $-\text{Im}(\chi^{(0)})\neq 0$, shown in Fig. \ref{Fig:opt_trans}. The FA particle-hole continuum extends linearly from in the $(q_y>0,\omega>0)$-plane. A second particle-hole spectrum is delimited from below by $\omega > \sqrt{4 e_0^2 + v^2 q_y^2}$ due to interband excitations involving the VP conduction and valence bands $n=\pm 1$, respectively. However, it vanishes at small momenta because the eigenstates associated with the VP conduction and valence bands are orthogonal at $\vect{q}=0$.

As we increase $\mu$ above the VP conduction band [see Figs. \ref{Fig:qz=0}(c), (d) and \ref{Fig:qz_nonzero}(c), (d)], the poles of the FA excitations remain unchanged because of the linear FA dispersion whereas that of the interband particle-hole continua gets heavily modified due to Pauli blocking at the conduction band minima. At low frequencies, intraband excitations of the $n=+1$ VP band induce a third particle-hole continuum. With $\mu>e_0$ just above the conduction band minimum, the VP band is approximately parabolic, and its quasi-1D character is apparent in the form of the particle-hole intraband spectrum with its typical exclusion dome for $q\ell_S < 2k_F$ with $k_F$ defined as $\mu = \sqrt{v^2 k_F^2  + e_0^2}$. 

In the $q_z = 0$ limit, two plasmon modes are present for $\mu < e_0$ as we can see in Figs. \ref{Fig:qz=0}(a), (b). The first one is the linearly dispersing FA plasmon with a gap at $q_y = 0$, in agreement with theoretical approaches using classical electrodynamics \cite{song_prb2017}, hydrodynamic description \cite{gorbar_prb2019}, or quantum-mechanical calculations \cite{andolina_prb2018,adinehvand_prb2019,chen_prb2019,ghosh_prb2020}. From the zeros of the real part of the equation $\epsilon^{\text{RPA}}(q_y,\omega) = 0$, we find the FA-plasmon disersion
\begin{align}
\label{eq:fa_plasmon_qz_zero}
    \omega \approx \text{sgn}(q_y) \frac{k_0 e^2}{4\pi^2 \epsilon_0 \epsilon_r} +  \left( 1 + \frac{2 k_F}{\sqrt{k_F^2 + \frac{2}{\ell_S^2}}}\delta_{n_F,1} \right) v q_y,
\end{align}
where $2k_0 = 2\sqrt{2m\Delta}$ is the separation between two Weyl nodes in the bulk, $\epsilon_0$ is the vacuum and $\epsilon_r$ the relative permittivity, while $n_F$ is the integer part of the ratio between $\mu$ and $e_0$. Let us first focus on the case where $n_F=0$. For positive $\omega$, the FA plasmon is allowed to propagate only in the direction of positive $q_y$, due to the chirality of the FA. For the usual Coulomb potential, recall that the 1D and 2D plasmon dispersions are linear and square-root, respectively. In spite of the the quasi-1D nature of the FA, the Coulomb potential remains 2D here, and one might naively expect a square-root plasmon dispersion. Surprisingly, this is not the case, and one finds a \textit{linear gapped} plasmon mode thanks to its chiral nature. As simple it is, Eq. \eqref{eq:fa_plasmon_qz_zero} accurately describes the mode found numerically in Fig. \ref{Fig:qz=0}(a), (b), even when the $n=\pm 1$ VP states are retained in the calculation. We emphasize that the experimentally measurable FA plasmon gap,
\begin{align}
\label{eq:gap}
    \gamma = \frac{k_0 e^2}{4\pi^2 \epsilon_0 \epsilon_r},
\end{align}
yields direct information about the separation $2k_0$ of the bulk Weyl nodes.

VP intraband excitations significantly modify the dispersion of the FA plasmon when $n_F=1$. The dispersion of FA plasmon at large $q_y$ remains almost linear with the same velocity $v$. At small $q_y$, although the gap sticks to the same value $\gamma$ in Eq. \eqref{eq:gap}, the FA plasmon acquires an enhanced velocity that can be further boosted by increasing the chemical potential as shown in Fig. \ref{Fig:qz=0}(c),(d). However, the change of velocity for finite $q_y$, as seen in Fig. \ref{Fig:qz=0}, is not captured by Eq. \eqref{eq:fa_plasmon_qz_zero}.

Since the bands are effectively 1D, the particle-hole continua are independent of $q_z$ if the coupling between FA and VP bands is omitted, as shown in Fig. \ref{Fig:qz_nonzero}. It is indeed legitimate to neglect this coupling in the long-wavelength limit where it scales as $\sim (q_z\ell_S)^2$. However, the plasmon dispersion gets strongly modified for $q_y < q_z$ due the $q_z$-dependence of the Coulomb interaction. Neglecting a small hybridization between the VP bands and the FA \cite{supp}, the FA-plasmon gap at small momenta becomes   
\begin{align}
\label{eq:fa_plasmon_qz_nonzero}
    \gamma' \approx \gamma \frac{q_y}{\sqrt{q_y^2 + q_z^2}}.
\end{align}

When $q_z = 0$, Eqs. \eqref{eq:fa_plasmon_qz_nonzero} and \eqref{eq:gap} coincide, and the FA plasmon is gapped as shown earlier. However when $q_z \ne 0$, the gap vanishes at $q_y=0$ as a consequence of the strong anisotropy of the FA state, which is only quasi-1D but embedded in a 2D manifold. When $q_y \gg q_z$, the FA plasmon disperses again linearly with slope $v$. This is further validated by our numerical calculations [see Fig. \ref{Fig:qz_nonzero}]: the FA plasmon gap vanishes when $\textbf{q} = q_z \hat{z}$. This singular behaviour of the gap at $\vect{q}=0$ is also reported in Refs.  \cite{song_prb2017,andolina_prb2018,adinehvand_prb2019,chen_prb2019,gorbar_prb2019,ghosh_prb2020}.

The second plasmon mode in Fig.\ref{Fig:qz=0}(a), (b) is the VP interband plasmon, which stems mostly from the $n=\pm 1$ interband excitations. It is also gapped and starts at a finite momentum for the same orthogonality reason that makes the spectral weight of the particle-hole continuum vanishingly small at $q_y\sim 0$, which makes sustained plasmonic osciallations impossible. The interband VP plasmon mode lies in the VP interband particle-hole region and is thus Landau-damped. However, since the amplitude of $-\text{Im}(\chi^{(0)})$ drops at high energy, this plasmon may be visible as an additional bump in Electron Energy Loss Spectroscopy (EELS), as we show later.  

It is interesting to point out that the spectrum in Fig. \ref{Fig:qz=0}(b), where $\mu=0.9 e_0$, is exactly the same as that of Fig. \ref{Fig:qz=0}(a), where the chemical potential is very close to the charge neutrality point. Indeed the interband excitations between the VP bands are unchanged as long as the chemical potential remains between the two VP bands. 

A third plasmon mode emerges when $\mu > e_0$ [see Figs. \ref{Fig:qz=0}(c), (d) and \ref{Fig:qz_nonzero}(c), (d)]. Interestingly, this mode exists in a region delimited by the particle-hole continua of the FA and the VP conduction bands. It starts at small but finite momentum and its energy disperses along with the upper boundary of the intraband continuum and eventually gets merged in it at larger momentum. One may naively think that this VP intraband plasmon originates only from intraband band excitations and has a square-root dispersion at small momenta \cite{supp}. However, our numerical calculations invalidate this picture, and one needs to take into account the other particle-hole continua, namely the linear one associated with the FA, which prohibits such a square-root dependence of an undamped plasmon. Moreover, remote VP interband excitations do not only modify $\epsilon_r$ in the low-energy modes because of the diverging density of states when the chemical potential crosses a VP conduction band. This modifies significantly dynamical screening and, as shown in Fig. \ref{Fig:qz=0}, the VP intraband plasmon acquires positive energy only at non-zero finite momentum and disperses linearly with a velocity smaller than $v$. Increasing $\mu$ from $1.01$ to $1.3 e_0$, the exclusion dome at low frequencies becomes wider. The available phase space for the VP intraband plasmon between the FA and the VP intraband continua has reduced even further so that this plasmon might be less visible at larger values of $\mu$.

\begin{center}
\begin{figure}
  \includegraphics[width=0.48\textwidth]{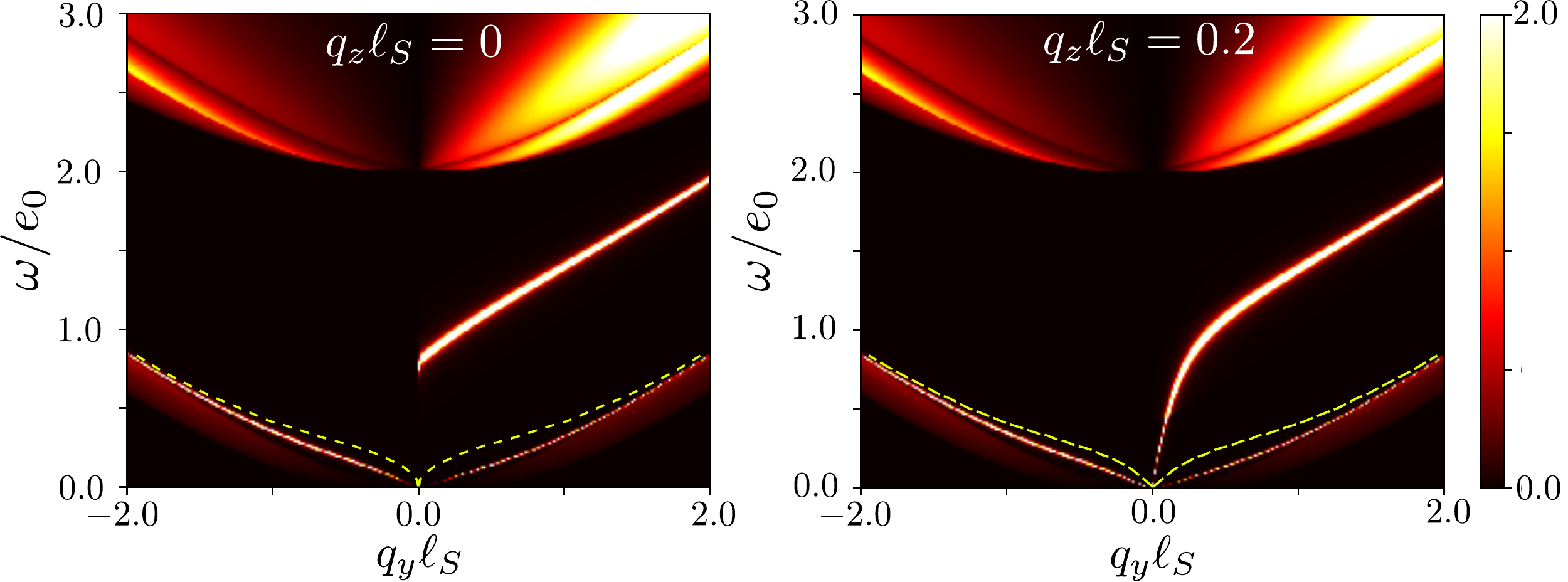}
    \caption{Electron loss function at $\mu = 1.01 e_0$ for $q_y \ell_S \in [-2.0,2.0]$: left for $q_z=0$ and right for $q_z \ell_S=0.2$. Yellow dashed lines show the symmetric VP intraband plasmon in the absence of the FA state. Three plasmon modes are all non-reciprocal and visible.}
  \label{Fig:qy_neg}
\end{figure}
\end{center}

\emph{Non-reciprocity.}---
To show what one can see in experiments, we plot in Fig. \ref{Fig:qy_neg} electron loss function $-\Im[1/\epsilon^{\text{RPA}}]$, measurable by EELS, in $(q_y, \omega>0)$-plane with intensity indicated by colorbar. One of the intriguing properties of FA is the non-reciprocity of FA plasmon, reflecting the chiral nature of FA state. Therefore, we should also study $-\Im[1/\epsilon^{\text{RPA}}]$ for $(q_y<0, \omega>0)$. The result for $\omega<0$ can be found easily by reversing simultaneously the sign of $\omega$ and $\vect{q}$ in known results. As shown in Fig. \ref{Fig:qy_neg} where $\mu = 1.01 e_0$, the FA plasmon is completely absent when $q_y < 0$ as well as the corresponding particle-hole continuum. Being non-reciprocal, FA plasmon only propagates in one direction with fixed velocity, highly desirable for applications. Strikingly, also the VP intraband plasmon is non-reciprocal even if it involves the $k_y\leftrightarrow -k_y$ symmetry of the $n=\pm$ VP bands [see Eq. (\ref{eq:energy_dispersion_n})]: it has a different dispersion for $q_y < 0$, which can be calculated analytically there
\begin{align}
    \label{eq:vp_intraband_plasmon_neg_qy}
    \omega \approx \text{sgn}(-q_y) v \left( |q_y| + \frac{2 k_F}{\sqrt{k_F^2 + \frac{2n}{\ell_S^2}}} \sqrt{q_y^2 + q_z^2}\right).
\end{align}
Contrary to $q_y>0$, it starts from the origin of $(q_y,\omega)$ and disperses with a velocity larger than $v$. This non-reciprocity is a consequence of the hybridization with the FA mode and particle-hole continuum, which is in close vicinity of the intraband VP plasmon for $q_y>0$ but further well separated in energy for $q_y<0$ \cite{supp}. The chirality of the FA modes thus induces a non-reciprocity in the other excitations due to their mutual coupling. This can also be seen in the VP interband plasmon, where the starting point moves to higher energies and larger momenta. As anticipated above, the VP interband plasmon is submerged amid particle-hole continuum but nevertheless visible on EELS.

\emph{Discussions.}---
We have investigated the effect of surface smoothness on the charge oscillation spectrum of a WSM surface. Within RPA calculations, we observe the emergence of two collective modes stabilized by the inter and intra VP band excitations, in addition to the FA plasmon. The plasmons exhibit anisotropy and non-reciprocity inherited from the underlying surface model. Our findings could be verified experimentally, e.g. in EELS, which in addition to a proof of these plasmons could probe the chirality of the FA. Furthermore,the plasmon gap in Eq. \eqref{eq:gap} gives us a direct experimental measure of the separation between the Weyl nodes.

We acknowledge financial support from Agence  Nationale de la Recherche (ANR project ``Dirac3D'') under Grant No. ANR-17-CE30-0023.

\bibliography{references}


\pagebreak
\begin{widetext}

\begin{center}
\textbf{\large Supplemental Materials for ``Surface plasmonics of Weyl semimetals"}
\end{center}
\setcounter{equation}{0}
\setcounter{figure}{0}
\setcounter{table}{0}
\setcounter{page}{1}
\makeatletter
\renewcommand{\theequation}{S\arabic{equation}}
\renewcommand{\thefigure}{S\arabic{figure}}
\renewcommand{\bibnumfmt}[1]{[S#1]}
\renewcommand{\citenumfont}[1]{S#1}

\section{S1. Solution of the Hamiltonian for the smooth surface of a time-reversal-broken Weyl semimetal}

In all calculations below, we use $\hbar = 1$ for notational simplicity. The basic Hamiltonian for an interface between a Weyl semimetal (WSM) and a (trivial) band insulator (``vacuum'')
\begin{align}
    H = v(k_x \sigma_x + k_y \sigma_y) + \left( \frac{k_z^2}{2m} - \Delta + \frac{2\Delta}{\ell}x \right) \sigma_z
\end{align}

reads, after a unitary transformation, 
    \begin{gather}
     	H_T = v 
        \begin{bmatrix}
        	-k_y & \frac{\sqrt{2}}{\ell_S} \hat{a}   \\
            \frac{\sqrt{2}}{\ell_S} \hat{a}^{\dagger} & k_y 
        \end{bmatrix}
    \end{gather}
where
\begin{align}
    \ell_S &= \sqrt{\ell \frac{v}{2\Delta}} \\
    \hat{a}^\dagger &= \frac{\ell_S}{\sqrt{2}} \left(k_x + i \frac{x - \langle x\rangle}{\ell_S^2}\right)  ,
\end{align}
and $\langle x\rangle = (\Delta-k_z^2/2m)\ell/2\Delta$ determines the average position of the surface state. It is important to note that $\langle x\rangle$ depends on the surface momenta and we explicitly mark this dependence on the spinor components $\ket{n,k_z}$. The eigenstates are thus of the form as
    \begin{align}
        \ket{\psi^{\lambda}_n} & = \frac{1}{\sqrt{2}}
        \begin{pmatrix}
        	u_{n,\lambda}(k_y) \ket{n-1,k_z} \\
            \lambda v_{n,\lambda}(k_y) \ket{n,k_z}
        \end{pmatrix} = \frac{1}{\sqrt{2}}
        \begin{pmatrix}
    	\sqrt{1-\lambda\frac{k_y}{\sqrt{k_y^2 + \frac{2n}{\ell_S^2}}}} \ket{n-1,k_z} \\
            \lambda \sqrt{1+\lambda\frac{k_y}{\sqrt{k_y^2 + \frac{2n}{\ell_S^2}}}} \ket{n,k_z}
        \end{pmatrix}
        & \text{if $n \geq 1$ }\\
        \ket{\psi_0} & = 
         \begin{pmatrix}
         	0 \\
            \ket{0,k_z} 
        \end{pmatrix} 
        &  \text{if $n = 0$ }
    \end{align}
    where $\lambda = \pm 1$ is band index and $\ket{n,k_z}$ is the eigenstates of quantum harmonic oscillator defined by $\hat{a}^{\dagger}(k_z)$ and $\hat{a}(k_z)$. The energy is thus
    \begin{align}
        E_n^{\lambda} &= \lambda v \sqrt{k_y^2 + \frac{2n}{\ell_S^2}} & \text{if $n \geq 1$ }\\
        E_0 &=v k_y & \text{if $n = 0$ }
    \end{align}
    
    Note that the FA state ($n=0$) has linear dispersion on $k_y$ breaking the parity symmetry $k_y \to -k_y$, which justifies the name of chiral state, and the VP states preserve this symmetry. Although the eigenstates live on a two-dimensional manifold $(k_y,k_z)$, their energies disperse only in the $k_y$-direction, the direction perpendicular to the interface and the line connecting the Weyl nodes at $k_z^W=\pm k_0=\pm \sqrt{2\Delta m}$ in reciprocal space. The bands are thus quasi-1D and the density of states of the VP states diverges at the band extremum. The dependence of $k_z$ is only encoded in the eigenstates, more precisely the cyclotron center of $\ket{n}$. 

\section{S2. Non-interacting dynamical polarization}
In this section, we give the analytical expressions of the non-interacting charge susceptibility. In the multi-band model, the charge susceptibility $\chi^{(0)}_{1,2}$ is in general a tensor
\begin{align}
    \chi^{(0)}_{1,2}(\vect{q},\omega) = \sum_{1,2} \frac{1}{V} \sum_{\vect{k}} \frac{f_D (E_1 (\vect{k}) ) - f_D (E_2 (\vect{k+q}) )}{\omega + E_1 (\vect{k}) - E_2 (\vect{k+q}) + i \eta} \times |F_{1,2} (\vect{k},\vect{k+q})|^2
\end{align}
where the number indices are shorthand notation of all degrees of freedom, $j=(n_j,\lambda_j)$ except the momentum $\vect{k}$, i.e. $E_1$ is the energy of the band with band order $n_1$ and band index $\lambda_1$, and $F_{1,2}(\vect{k},\vect{k+q})$ is the overlap function 

\begin{align}
    |F_{1,2}(\vect{k},\vect{k+q})|^2 &= |\bracket{1,\vect{k}}{2,\vect{k+q}}|^2 \\
                                     &= \frac{1}{4} | u_1^*(k_y) u_2(k_y +q_y) R_{n_1-1,n_2-1}(k_z,k_z+q_z) + \lambda_1 \lambda_2 v_1^*(k_y) v_2(k_y +q_y) R_{n_1,n_2}(k_z,k_z+q_z) |^2
\end{align}
where
\begin{align}
R_{n_1,n_2}(k_z,k_z+q_z) &= \bracket{n_1, k_z}{n_2, k_z + q_z} \\
&= \sqrt{\frac{n_2 !}{n_1 !}}\alpha^{n_1-n_2} L_{n_2}^{(n_1-n_2)}(|\alpha|^2) e^{-\frac{|\alpha|^2}{2}} \quad \text{if $n_1 \ge n_2$} \\
&= \sqrt{\frac{n_1 !}{n_2 !}}\alpha^{n_2-n_1} L_{n_1}^{(n_2-n_1)}(|\alpha|^2) e^{-\frac{|\alpha|^2}{2}} \quad \text{if $n_2 > n_1$},
\end{align}
and
\begin{align}
    \alpha = \alpha(k_z,k_z+q_z) = i \sqrt{\frac{l}{8m^2 v(\Delta+\Delta^\prime)} }(2 k_z + q_z) q_z = i \sqrt{\frac{\Delta}{2(\Delta+\Delta^\prime)} \frac{\Delta l}{v} } \frac{(2 k_z + q_z) q_z}{k_0^2}.
\end{align}
This is precisely a manifestation of the strong anisotropy of the edge states in WSM, and we can already anticipate a highly anisotropic dynamical polarization. 

Even if the overlap function is complicated, the expressions can be simplified significantly in the long-wavelength limit. Let us consider $q_z/k_0 \to 0$, $k_0$ appears to be a cut-off in the long-wavelength limit. When $q_z = 0$, $R_{n_1,n_2}$ is simplified to $\delta_{n_1,n_2}$ due to the orthogonality of the wave functions. Therefore,
\begin{align}
    |F_{1,2}(\vect{k},\vect{k+q})|^2 = \frac{1}{4} \delta_{n_1,n_2} |u_1^* u_2 + \lambda_1 \lambda_2 v_1^* v_2|^2
\end{align}
and one notices that only excitations $n \to n$ are allowed. Suppose that $q_z \ne 0$ and for example $n_1 \ge n_2$
\begin{align}
    R_{n_1,n_2}(k_z,k_z+q_z) \sim \sqrt{\frac{n_2 !}{n_1 !}} \binom{n_2}{n_1} \alpha^{n_1-n_2} \propto q_z^{n_1-n_2}.
\end{align}
The $n \to n$ excitations are therfore still the leading contribution to the charge susceptibility, and the coupling strength of other excitations decays as $q_z^{|n_1-n_2|}$.

Based on the arguments exposed above, we consider in the rest of this section $q_z = 0$. Accordingly, the tensor $\chi^{(0)}_{1,2}$ is diagonal and can be treated as a scalar. We can thus divide it into contributions from different excitation invoking different bands,
\begin{align}
    \chi^{(0)} = \sum_{n \ge 1} \chi_{n}(q_y,\omega) + \chi_{0} (q_y, \omega).
\end{align}
For $n=0$,
\begin{align}
     \chi_{0}(q_y,\omega) =  \frac{1}{V} \sum_{\vect{k}} \frac{f_D (E_0 (\vect{k}) ) - f_D (E_0 (\vect{k+q}) )}{\omega + E_0 (\vect{k}) - E_0 (\vect{k+q}) + i \eta} ,
\end{align}
while for $n \ge 1$
\begin{align}
    \chi_{n}(q_y,\omega) = \sum_{\lambda_1,\lambda_2} \frac{1}{V} \sum_{\vect{k}} \frac{f_D (E_n^{\lambda_1} (\vect{k}) ) - f_D (E_n^{\lambda_2} (\vect{k+q}) )}{\omega + E_n^{\lambda_1} (\vect{k}) - E_n^{\lambda_2} (\vect{k+q}) + i \eta} \times |F_{\lambda_1,\lambda_2}^{n} (\vect{k},\vect{k+q})|^2,
\end{align}
where
\begin{align}
    |F_{\lambda_1,\lambda_2}^{n} (\vect{k},\vect{k+q})|^2 
    = \frac{1}{2} \left( 1 + \lambda_1 \lambda_2 \frac{k_y(k_y + q_y) + \frac{2n}{\ell_S^2}}{\sqrt{\left( k_y^2 + \frac{2n}{\ell_S^2} \right) \left((k_y+q_y)^2 + \frac{2n}{\ell_S^2}\right)} } \right).
\end{align}

To obtain the plasmon mode, we need first calculate the imaginary part of each susceptibility for $n=0$ and $n\geq 1$ from which their real part can then be deduced by the Kramers-Kronig relations. The single-pair excitations (SPE) resides in the domain in the $(q_y,\omega)$-space where the imaginary part is not vanishing. We assume $T=0$ in the following calculations and we can consider only $\omega>0$. 

\subsection{Fermi arc $n=0$}
Because of the chirality of the FA state, particle-hole excitations with $\omega > 0$ require $q_y > 0$,
\begin{align}
    \chi_{0}(q_y>0,\omega>0) &= \frac{k_0}{2 \pi^2} \frac{q_y}{\omega - v q_y + i \eta } \\
    &= \frac{k_0 q_y}{2 \pi^2} \left( \frac{\omega-v q_y}{(\omega-v q_y)^2 + \eta^2} - i\frac{\eta}{(\omega-v q_y)^2 + \eta^2}\right) .
\end{align}
Note that this expression is valid for $\omega$ and $q_y$ of arbitrary sign and that $\chi_0$ is independent of the chemical potential due to the linear dispersion of the FA state and a well-defined SPE $\omega=v q_y$. Using RPA and $V(q) = e^2/2 \epsilon_0 q$ which we will justify in the following section, we can easily find the FA plasmon
\begin{align}
    \omega = v q_y + \frac{k_0 e^2}{4\pi^2 \epsilon_0} \text{sgn}(q_y),
\end{align}
which is gapped and disperses linearly when $q_z = 0$.

\subsection{Massive states $n \ge 1$}
The VP states are not chiral such that the sign of $q_y$ is irrelevant.
\subsubsection{Charge neutral point}
When the chemical potential is at the charge neutral point, only the interband excitations are possible. Since the VP states are gapped, the following results are also valid for finite chemical potential if only it resides between two band extrema of the VP states. The imaginary part of $\chi_n$ reads
\begin{align}
    \Im[\chi_n^-] (q_y,\omega>0)=- \frac{k_0}{\pi v} \frac{\frac{4n}{\ell_S^2} q_y^2}{\left( \frac{\omega^2}{v^2} - q_y^2 \right)^2\sqrt{1-\frac{\frac{8n}{\ell_S^2}}{\frac{\omega^2}{v^2}-q_y^2} }} \Theta \left( \frac{\omega^2}{v^2} - q_y^2 - \frac{8n}{\ell_S^2} \right)
\end{align}
where the minus sign on $\chi_n$ refers to the contribution from the $\lambda=-$ valence band. The particle-hole spectrum $\omega > v\sqrt{q_y^2 + \frac{8n}{\ell_S^2}}$ is above the FA plasmon for realistic value of $k_0$. Note that the imaginary part is proportional to $q_y^2$ and vanishing at $q_y = 0$ due to the vanishing overlap function when $q_y = 0$. This is simply because the eigenstates of Bloch Hamiltonian for a given $k$ are orthogonal. The corresponding real part reads

\begin{align}
    \Re[\chi_n^-] &= - \frac{4 k_0 \frac{2n}{\ell_S^2} q_y^2}{\pi^2 v} \left( \frac{1}{\frac{4n}{\ell_S^2} \left(q_y^2 - \frac{\omega^2}{v^2} \right)} + \frac{\pi - 2\arctan A}{\left|q_y^2 - \frac{\omega^2}{v^2} \right|^{\frac{3}{2}} \left|\frac{8n}{\ell_S^2} + q_y^2 - \frac{\omega^2}{v^2} \right|^{\frac{1}{2}}} \right) \quad \text{if $q_y^2 \le \frac{\omega^2}{v^2} \le  q_y^2 + \frac{4n}{\ell_S^2}$} \\
    &= - \frac{4 k_0 \frac{2n}{\ell_S^2} q_y^2}{\pi^2 v} \left( \frac{1}{\frac{4n}{\ell_S^2} \left(q_y^2 - \frac{\omega^2}{v^2} \right)} - \frac{\log|\frac{A+1}{A-1}|}{\left|q_y^2 - \frac{\omega^2}{v^2} \right|^{\frac{3}{2}} \left|\frac{8n}{\ell_S^2} + q_y^2 - \frac{\omega^2}{v^2} \right|^{\frac{1}{2}} } \right) \quad \text{otherwise}
\end{align}
where
\begin{align}
    A = \sqrt{ \left| \frac{\frac{8n}{\ell_S^2} + q_y^2 - \frac{\omega^2}{v^2}}{ q_y^2 - \frac{\omega^2}{v^2}} \right| }.
\end{align}
Using RPA and $V(q) = e^2/2\epsilon_0 q$, we can prove the existence of a plasmon mode damped by the interband SPE. Let's check two limits. When $\omega^2/v^2 - q_y^2 \gg 8n/\ell_S^2$,
\begin{align}
    1 - V(q_y) \Re[\chi_n^-](\omega, q_y) \to 1^{-}.
\end{align}
When $\omega^2/v^2 - q_y^2 \to^+ 8n/\ell_S^2$,
\begin{align}
    1 - V(q_y) \Re[\chi_n^-](\omega, q_y) &= 1- \frac{e^2}{2\pi^2 \epsilon_0 v} \frac{k_0 |q_y|}{\frac{2n}{\ell_S^2}}, 
\end{align}
which can only be satisfied for non-zero values of $|q_y|$.
This plasmon emerges only at finite $|q_y|$ for the same reason as the vanishing imaginary part.

\subsubsection{Band-crossing chemical potential}
Now we consider the case of a chemical potential crossing one of the VP bands. Since VP states are particle-hole symmetric, we consider a chemical potential in the conduction band ($n=+1$),
\begin{align}
    \mu = v\sqrt{k_F^2 + \frac{2n}{\ell_S^2}} > 0,
\end{align}
where we define a Fermi wavevector $k_F$. Now intraband excitations in the $\lambda=+$ conduction band are possible and we have to suppress a part of interband contributions in the susceptibility due to Pauli blocking. Since the latter does not change the boundary of the particle-hole spectrum (checked by numerical calculations), we give here only the expressions for the intraband excitations. The imaginary part for intraband excitations reads
\begin{align}
    \Im[\chi_{n,\text{intra}}^+] (q_y,\omega>0) &= - \frac{k_0}{4 \pi v } \frac{q_y^2 \frac{8n}{\ell_S^2}}{\left|q_y^2 - \frac{\omega^2}{v^2} \right|^{\frac{3}{2}} \left|\frac{8n}{\ell_S^2} + q_y^2 - \frac{\omega^2}{v^2} \right|^{\frac{1}{2}}} \\
    & \times \left( \Theta \left(2 k_F - \left| q_y - \frac{\omega}{v} \sqrt{1+\frac{\frac{8n}{\ell_S^2}}{q_y^2-\frac{\omega^2}{v^2}}} \right| \right) - \Theta \left(2 k_F - \left| q_y + \frac{\omega}{v} \sqrt{1+\frac{\frac{8n}{\ell_S^2}}{q_y^2-\frac{\omega^2}{v^2}}} \right| \right)\right).
\end{align}
For the intraband excitations in the $n=+1$ VP band, we find a particle-hole region between two parallel parabolic-like boundaries excluding a dome from 0 to $2k_F$, as it should be for the Lindhard function for a one-dimensional parabolic band, as shown in Figure \ref{fig:ana_ph} where we gather also the other excitations (intra-FA and inter-VP band excitations from $-1$ to $+1$). This is due to the one-dimensional dispersion of the massive VP states. We can anticipate that when the chemical potential is high the Dirac linear dispersion of the VP states is probed. So the two parallel parabolic-like boundaries becomes linear and parallel to the line defined by $\omega = v q_y$. 

\begin{figure}[h]
    \centering
    \includegraphics[width=0.6 \textwidth]{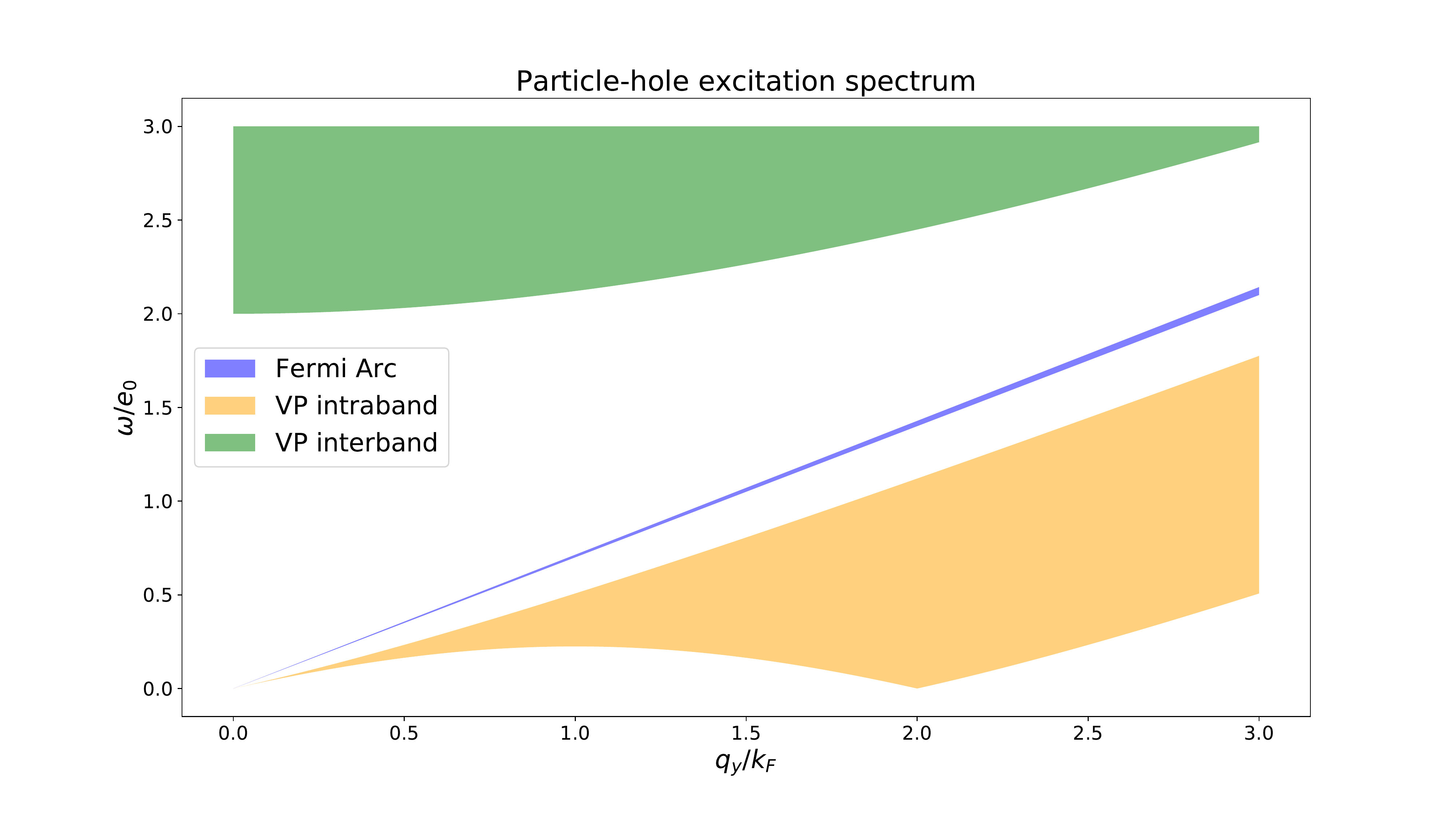}
    \caption{Particle-hole region}
    \label{fig:ana_ph}
\end{figure}

The Kramers-Konig relations yield a real part of the intraband contributions that reads
\begin{align}
\nonumber
    \Re[\chi_{n,\text{intra}}^+] &= \frac{k_0 \frac{2n}{\ell_S^2} q_y^2}{\pi^2 v} \left( g_1\left( q_y^2 - \left| \sqrt{k_F^2 + \frac{2n}{\ell_S^2}} - \sqrt{|k_F-|q_y||^2 + \frac{2n}{\ell_S^2}} \right| \right) - g_1\left( q_y^2 - \left| \sqrt{k_F^2 + \frac{2n}{\ell_S^2}} - \sqrt{|k_F+|q_y||^2 + \frac{2n}{\ell_S^2}} \right| \right) \right) \\ &\quad \text{if $q_y^2 \le \frac{\omega^2}{v^2} \le  q_y^2 + \frac{4n}{\ell_S^2}$}\\
    \nonumber
    &= \frac{k_0 \frac{2n}{\ell_S^2} q_y^2}{\pi^2 v} \left( g_2\left( q_y^2 - \left| \sqrt{k_F^2 + \frac{2n}{\ell_S^2}} - \sqrt{|k_F-|q_y||^2 + \frac{2n}{\ell_S^2}} \right| \right) - g_2\left( q_y^2 - \left| \sqrt{k_F^2 + \frac{2n}{\ell_S^2}} - \sqrt{|k_F+|q_y||^2 + \frac{2n}{\ell_S^2}} \right| \right) \right) \\
    &\quad \text{otherwise}
\end{align}
where
\begin{align}
    g_1 (x) &= \frac{1}{\frac{4n}{\ell_S^2} \left(q_y^2 - \frac{\omega^2}{v^2} \right)} \sqrt{\frac{x+\frac{8n}{\ell_S^2}}{x}} - \frac{2\arctan f(x)}{\left|q_y^2 - \frac{\omega^2}{v^2} \right|^{\frac{3}{2}} \left|\frac{8n}{\ell_S^2} + q_y^2 - \frac{\omega^2}{v^2} \right|^{\frac{1}{2}}}\\
    g_2 (x) &= \frac{1}{\frac{4n}{\ell_S^2} \left(q_y^2 - \frac{\omega^2}{v^2} \right)} \sqrt{\frac{x+\frac{8n}{\ell_S^2}}{x}}  - \frac{\log|\frac{f(x)+1}{f(x)-1}|}{\left|q_y^2 - \frac{\omega^2}{v^2} \right|^{\frac{3}{2}} \left|\frac{8n}{\ell_S^2} + q_y^2 - \frac{\omega^2}{v^2} \right|^{\frac{1}{2}} }\\
    f(x) &= \sqrt{ \left| \frac{\frac{8n}{\ell_S^2} + q_y^2 - \frac{\omega^2}{v^2}}{ q_y^2 - \frac{\omega^2}{v^2}} \right| \cdot \frac{x+\frac{8n}{\ell_S^2}}{x}}.
\end{align}
In the long wavelength limit ($q_y \to 0$), 
\begin{align}
    \Re[\chi_{n,\text{intra}}^+] = \frac{k_0}{\pi^2 v} \frac{k_F}{\sqrt{k_F^2 + \frac{2n}{\ell_S^2} }} \frac{v^2 q_y^2}{\omega^2 - v^2 q_y^2}.
\end{align}
Recall that for a spinless 1D parabolic band, these expressions may be simplified, and one obtains
\begin{align}
    \Re[\chi_{\text{para}}] = \frac{ k_F q_y^2}{\pi m \omega^2}.
\end{align}
At the band minimum where $k_F \to 0$, $\Re[\chi_{n,\text{intra}}^+]$ becomes indeed
\begin{align}
    \Re[\chi_{n,\text{intra}}^+] = \frac{2 k_0}{2\pi}\frac{k_F q_y^2}{\pi m_D \omega^2},
\end{align}
where $m_D$ is the Dirac mass $\sqrt{2}/v \ell_S$. The strong anisotropy is explicitly unveiled in this form: dispersed in the transverse direction and flat in the longitudinal direction, effectively 1D band. However, the prefactor $2k_0/2\pi$ reminds us that the underlying manifold is 2D. Indeed, $2k_0$ is the extension of the VP bands in the $k_z$-direction delimited by the two Weyl nodes at $k_z=\pm k_0$, and $2k_0/2\pi$ is thus its contribution to the density of states. It is also interesting to calculate the dynamical polarization for a 1D linear band $E = \pm v k_y$ for a positive chemical potential and the excitation momentum $|q_y| < k_F$, that is
\begin{align}
    \Re[\chi_{\text{lin}}] = \frac{k_0}{\pi^2 v} \frac{v^2 q_y^2}{\omega^2 - v^2 q_y^2},
\end{align}
which coincides with $\Re[\chi_{n,\text{intra}}^+]$ when $k_F \gg 2n/\ell_S^2$.

Using RPA and $V(q) = e^2 / 2 \epsilon_0 q$, a square-root intraband plasmon is obtained:
\begin{align}
    \omega = \sqrt{\frac{e^2 v k_0}{2 \pi^2 \epsilon_0} \frac{k_F}{\sqrt{k_F^2 + \frac{2n}{\ell_S^2} }} q_y + v^2 q_y^2}.
\end{align}
The intraband plasmon is pushed up with increasing chemical potential and $k_F$.

\section{S3. Quasi-two-dimensional random phase approximation}
In this section we show how to use the random-phase approximation (RPA) in our quasi-2D Hamiltonian. We show $\ell_S$ as another length scale cut-off to consider our Hamiltonian as a quasi-2D system. We have already shown the complicated overlap function when $q_z \ne 0$. So the matrix element of the Coulomb interaction must be at least as messy as the overlap function. We first show the matrix element of the Coulomb interaction in the most general form without any approximation. Then, we explicitly show how the long-wavelength approximation simplifies the calculations considerably.

\subsection{Coulomb interaction matrix element}
Consider the wavefunction of the eigenstates
 \begin{align}
        \bracket{\vect{r}}{\psi_{n,\lambda}} & = \frac{1}{\sqrt{S}} e^{i \vect{k_\parallel}\cdot\vect{r_\parallel}} \chi_{n,\lambda}(x,k_\parallel) \\
        &= \frac{1}{\sqrt{2S}} e^{i \vect{k_\parallel}\cdot\vect{r_\parallel}} 
        \begin{pmatrix}
        	u_{n,\lambda}(k_y) \bracket{x}{n-1}(k_z) \\
            \lambda v_{n,\lambda}(k_y) \bracket{x}{n}(k_z)
        \end{pmatrix},
\end{align}
where $S$ is the size of the surface and in terms of which the Coulomb interaction operator reads
\begin{align}
    \hat{V} = \frac{1}{2} \frac{e^2}{4 \pi \epsilon_0 S} \sum_{1,2,3,4} \sum_{k_1,k_2,q} \iint dx dx' \frac{2 \pi e^{-|q(x-x')|}}{|q|} \chi_{1}^\dagger(x,k_1) \chi_{3}(x,k_1 + q) \chi_{2}^\dagger(x',k_2) \chi_{4}(x',k_2 - q) \hat{c}^{\dagger}_{1,k_1} \hat{c}^{\dagger}_{2,k_2} \hat{c}_{4,k_2-q} \hat{c}_{2,k_1 + q}.
\end{align}
For smooth interfaces with $\ell \gg \ell_S$, one can show that the dominant contribution to the Coulomb operator comes from the region where $q(\langle x\rangle - \langle x'\rangle) \ll 1$, such that the effective Coulomb potential reads
\begin{align}
    \hat{V} \approx \frac{1}{2} \frac{e^2}{2 \epsilon_0 S |q|} \sum_{1,2,3,4} \sum_{k_1,k_2,q} F_{1,3} (k_1 , k_1 + q) F_{2,4} (k_2 , k_2 - q) \hat{c}^{\dagger}_{1,k_1} \hat{c}^{\dagger}_{2,k_2} \hat{c}_{4,k_2-q} \hat{c}_{2,k_1 + q}.
\end{align}
The complicated form of the overlap function renders further simplification impossible unless we suppose $q_z = 0$. Then we can define
\begin{align}
    \hat{V} \approx \frac{1}{2} \sum_{1,2,3,4} \sum_{q} V_{13,24}(q) \frac{1}{S}\sum_{k_1,k_2} F^{n_1}_{\lambda_1,\lambda_3} (k_1 , k_1 + q) F^{n_2}_{\lambda_2,\lambda_4} (k_2 , k_2 - q) \hat{c}^{\dagger}_{1,k_1} \hat{c}^{\dagger}_{2,k_2} \hat{c}_{4,k_2-q} \hat{c}_{2,k_1 + q},
\end{align}
where the diagonal two-dimensional Coulomb interaction matrix element is defined as
\begin{align}
    V_{13,24}(q) = \frac{e^2}{2 \epsilon_0 |q|} \delta_{n_1,n_3} \delta_{n_2,n_4}.
\end{align}
Thus the interaction tensor becomes a scalar, and the RPA polarizability reads
\begin{align}
    \chi(q_y,\omega) = \frac{\chi^{(0)} (q_y,\omega)}{ 1- V_{\text{2D}}(|q_y|) \chi^{(0)}(q_y,\omega)},
\end{align}
where
\begin{align}
    \chi^{(0)} &= \sum_{n \ge 1} \chi_{n}(q_y,\omega) + \chi_{0} (q_y, \omega) \\ \text{and}\quad
    V_{\text{2D}}(|q_y|) &= \frac{e^2}{2 \epsilon_0 |q_y|}
\end{align}

When $q_z \ne 0 $, we can in general not simplify the Coulomb interaction operator to the previous form. Luckily, when we consider only a few bands, the long-wavelength limit renders the expressions tractable. 

\subsection{Three-band model}
Let us consider a three-band model with the chiral FA state and the two $n=\pm 1$ VP states. As simple as it is, this toy model gives already many promising results on plasmon. As shown before, it is difficult to deal with overlap functions when $q_z \ne 0$ because of $R_{n_1,n_2} \ne 0$ for $n_1 \ne n_2$. In the long wavelength limit, i.e. $q/k_0 \ll 1$, we can write them explicitly for $n_1, n_2 \in \{ 0,1 \}$ as
\begin{align}
    R_{0,0}(k_z,k_z + q_z) &= R_{1,1}(k_z,k_z + q_z) = 1 + O(q^2) \approx 1\\
    R_{0,1}(k_z,k_z + q_z) &= R_{1,0}(k_z,k_z + q_z) = \alpha(k_z,k_z + q_z) + O(q^3) \approx |\alpha|,
\end{align}
where $\alpha \sim q$. Then we can rewrite overlap functions as 
\begin{align}
    |F_{1,2}(\vect{k},\vect{k+q})|^2 = \frac{1}{4} |u_1^* u_2 + \lambda_1 \lambda_2 v_1^* v_2|^2 |R_{n_1,n_2}(k_z,k_z + q_z)|^2 = |G_{1,2}(k_y,k_y + q_y)|^2 |R_{n_1,n_2}(k_z,k_z + q_z)|^2,
\end{align}
where we decouple the $q_y$ and $q_z$ directions. On the RPA level, the final expression for $\chi$ reads
\begin{align}
    \chi(q,\omega) = \frac{\chi^{(0)} (q,\omega)}{ 1- V_{\text{2D}}(|q|) \chi^{(0)}(q,\omega)},
\end{align}
where
\begin{align}
    \chi^{(0)} &= \sum_{1,2} \chi^0_{1,2}(q,\omega), \\
    \chi^{(0)}_{1,2}(q,\omega) &= \sum_{1,2} \frac{1}{V} \sum_{\vect{k}} \frac{f_D (E_1 (k_y) ) - f_D (E_2 (k_y + q_y) )}{\omega + E_1 (k_y) - E_2 (k_y+q_y) + i \eta} \times |G_{1,2}(k_y,k_y + q_y)|^2 |R_{n_1,n_2}(k_z,k_z + q_z)|^2, \\ \text{and} \quad
    V_{\text{2D}}(|q|) &= \frac{e^2}{2 \epsilon_0 |q|}.
\end{align}

Before we calculate the full dynamical polarization including all the contributions in the three-band model, let us write the dispersion for FA plasmon and VP intraband plasmon when $q_z \ne 0$. In the long-wavelength limit, the coupling between states of different $n$, for example $n = 0$ and $n = 1$, is negligible because it is at least proportional to $q_z^2$. Thus, we simply need to replace $q_y$ by $q \cos \theta$ with $\cos \theta = q_y / \sqrt{q_y^2 + q_z^2}$ in the previous expressions of dynamical polarization. For the FA state,
\begin{align}
    \Re[\chi_{0}] &= \frac{k_0}{2 \pi^2} \frac{q \cos \theta}{\omega - v q \cos \theta},
\end{align}
and the FA plasmon dispersion reads
\begin{align}
    \omega = v q \cos \theta + \frac{k_0 e^2}{4 \pi^2 \epsilon_0} \cos \theta.
\end{align}
For the $n=1$ VP state,
\begin{align}
    \Re[\chi_{n,\text{intra}}^+] = \frac{k_0}{\pi^2 v} \frac{k_F}{\sqrt{k_F^2 + \frac{2n}{\ell_S^2} }} \frac{v^2 q^2 \cos^2 \theta}{\omega^2},
\end{align}
and the VP intraband plasmon dispersion reads
\begin{align}
    \omega = \cos \theta \sqrt{\frac{e^2 v k_0}{2 \pi^2 \epsilon_0} \frac{k_F}{\sqrt{k_F^2 + \frac{2n}{\ell_S^2} }} q_y}.
\end{align}
As shown in the main text of the paper, this expression cannot explain the linear dispersion of the VP intraband plasmon, and we need to take into account the influence of the FA state.

In the long-wavelength limit, assuming $q_z = 0$, we have
\begin{align}
    \Re[\chi_{\text{tot}}] = \frac{k_0}{\pi^2 v} \left( \frac{k_F}{\sqrt{k_F^2 + \frac{2n}{\ell_S^2}}} \cdot \frac{q_y^2}{\frac{\omega^2}{v^2} - q_y^2} + \frac{1}{2} \cdot \frac{q_y}{\frac{\omega}{v} - q_y} \right).
\end{align}
The first term in the parenthesis comes from the VP intraband excitations and the second one from the intra-FA excitations. Here we do not include $\Re[\chi_n^-]$ since it describes an insulator playing the role of substrate, which only change $\epsilon_0$ to $\epsilon_0 \epsilon_r$. For now, we do not take into the interband part in $\Re[\chi_n^+]$ either. As we will show later, this is essential for VP intraband plasmon. Using RPA and $V(q_y) = e^2 / 2 \epsilon_0 |q_y|$, the dielectric function reads
\begin{align}
    \epsilon = 1 - \frac{e^2 k_0}{2 \epsilon_0 \pi^2 v} \left( \frac{k_F}{\sqrt{k_F^2 + \frac{2n}{\ell_S^2}}} \cdot \frac{|q_y|}{\frac{\omega^2}{v^2} - q_y^2} + \frac{1}{2} \cdot \frac{\text{sgn}(q_y)}{\frac{\omega}{v} - q_y} \right).
\end{align}
To find the plasmon mode, we need to solve the following equation
\begin{align}
    \omega^2 - \text{sgn}(q_y) \gamma \omega - \gamma v |q_y| \left( 1 + \frac{2 k_F}{\sqrt{k_F^2 + \frac{2n}{\ell_S^2}}} \right) - v^2 |q_y|^2 = 0
\end{align}
where $\gamma = e^2 k_0 / 4 \epsilon_0 \pi^2$ is the gap of FA plasmon. This equation is readily solved, and one finds
\begin{align}
    \omega = \frac{\text{sgn}(q_y) \gamma \pm \sqrt{\gamma^2 + 4 \gamma v |q_y| \left( 1 + \frac{2 k_F}{\sqrt{k_F^2 + \frac{2n}{\ell_S^2}}} \right) + 4 v^2 |q_y|^2}}{2}.
\end{align}
As $q_y$ is small, we have
\begin{align}
    \omega = \text{sgn}(q_y) \gamma +  \left( 1 + \frac{2 k_F}{\sqrt{k_F^2 + \frac{2n}{\ell_S^2}}} \right) v q_y \quad \text{or} \quad  - \left( 1 + \frac{2 k_F}{\sqrt{k_F^2 + \frac{2n}{\ell_S^2}}} \right) v q_y.
\end{align}
Most saliently, the VP intraband plasmon has linear behavior compared to a square-root-like dispersion expected naively in our previous calculations, where we consider only the VP intraband excitations without taking into account the FA plasmon. For $q_z \ne 0$, we can find the plasmon
\begin{align}
    \omega = \frac{\gamma \cos \theta \pm  \gamma |\cos \theta|}{2} \pm v q \left( |\cos \theta| + \frac{2 k_F}{\sqrt{k_F^2 + \frac{2n}{\ell_S^2}}} \right)
\end{align}
where we omit the coupling between FA and VP band which is only a correction of $q_z^2$.

This result can explain the enhanced slope in the dispersion of the FA plasmon at small $|q_y|$ and the VP intraband plasmon when $q_y < 0$. However, we do not find an intraband plasmon for $q_y > 0 $, which we observe in our numerical calculations. Indeed, as we have anticipated, the interband part in $\Re[\chi_{n,inter}^+]$ accounts for this plasmon,
\begin{align}
    \Re[\chi_{n,inter}^+] &= \frac{1}{V} \sum_{\vect{k}} \frac{f_D (E_n^{+} (\vect{k}) )}{\omega + E_n^{+} (\vect{k}) - E_n^{-} (\vect{k+q}) + i \eta} \times |F_{+,-}^{n} (\vect{k},\vect{k+q})|^2 \\
    &+ \frac{1}{V} \sum_{\vect{k}} \frac{- f_D (E_n^{+} (\vect{k+q}) )}{\omega + E_n^{-} (\vect{k}) - E_n^{+} (\vect{k+q}) + i \eta} \times |F_{-,+}^{n} (\vect{k},\vect{k+q})|^2.
\end{align}
To simplify the discussion, let us assume that $k_F \gg \sqrt{2n}/\ell_S$ so that the energy dispersion is $\omega = \pm v |q_y|$. Also, $k_F / \sqrt{k_F^2 + 2n/\ell_S^2} \approx 1$. Neglecting the overlap function which is only a correction of $q_y^2$, we have
\begin{align}
    \Re[\chi_{n,inter}^+] &= \frac{k_0}{2 \pi^2 v} \int_{-k_F}^{k_F} d k_y \left( \frac{1}{\frac{\omega}{v} + |k_y| + |k_y + q_y| } - \frac{1}{\frac{\omega}{v} - |k_y| - |k_y - q_y| } \right) \\
    & = \frac{k_0}{2 \pi^2 v} \left( \log \left( \frac{4v^2 k_F^2}{v^2 q_y^2 - \omega^2} \right)  -\frac{2 q_y^2}{\frac{\omega^2}{v^2}-q_y^2} \right).
\end{align}
where $q_y>0$. Then,
\begin{align}
    \Re[\chi_n^+] = \Re[\chi_{n,inter}^+] + \Re[\chi_{n,intra}^+] = \frac{k_0}{2 \pi^2 v} \log\left( \frac{4 v^2 k_F^2}{v^2 q_y^2 - \omega^2}\right) 
\end{align}
where we set $0< \omega < v q_y$ as we have observed numerically. Therefore the RPA dielectric function reads
\begin{align}
    \epsilon = 1 - \frac{e^2 k_0}{4 \epsilon_0 \pi^2 v} \left( \frac{1}{|q_y|} \log\left( \frac{4 v^2 k_F^2}{v^2 q_y^2 - \omega^2}\right)  +  \frac{\text{sgn}(q_y)}{\frac{\omega}{v} - q_y} \right).
\end{align}
This is a transcendental equation that yields a plasmon but cannot be solved analytically. If we omit the second term in the parenthesis, we find
\begin{align}
    \label{eq:vp_intraband_qz_zero}
    \omega = v \sqrt{\left( 1- \frac{2 v^2 k_F^2}{\gamma^2}\right) q^2 + 4k_F^2 \frac{ v}{\gamma} q - 4 k_F^2},
\end{align}
where we have a velocity smaller than $v$. Although this formula cannot quantitatively match our numerical calculations shown in the main text, it demonstrates qualitatively the importance of $\Re[\chi_{n,inter}^+]$.

\end{widetext}

\end{document}